\documentclass{mn2e}
\usepackage{amssymb,graphicx}

\newcommand{\eg}{\textit{e.g.}\ }
\newcommand{\ie}{\textit{i.e.}\ }

\newcommand{\Jvec}{{\bf J}}
\newcommand{\Bvec}{{\bf B}}

\newcommand{\Uvec}{{\bf U}}
\newcommand{\Evec}{{\bf E}}

\newcommand{\Igun}{I_{\rm gun}}
\newcommand{\Vgun}{V_{\rm gun}}

\newcommand{\psigun}{\psi_{\rm gun}}
\newcommand{\Ti}{T_{\rm i}}
\newcommand{\Te}{T_{\rm e}}

\newcommand{\nee}{n_{\rm e}}

\newcommand{\agun}{\alpha_{\rm gun}}

\newcommand{\Bt}{B_\theta}
\newcommand{\Bz}{B_{\rm Z}}

\newcommand{\Alfven}{Alfv\'{e}n }
\newcommand{\VA}{V_{\rm A}}
\newcommand{\rhoi}{\rho_{\rm i}}

\title{A Laboratory Plasma Experiment for Studying Magnetic Dynamics
of Accretion Discs and Jets}
\author[S. C. Hsu and P. M. Bellan]
       {S. C. Hsu and P. M. Bellan \\
        Dept.\ of Applied Physics, California Institute of Technology,
	Pasadena, CA\, 91125}
\date{}

\pagerange{\pageref{firstpage}--\pageref{lastpage}}
\pubyear{2002}

\begin{document}
\maketitle
\label{firstpage}

\begin{abstract}
This work describes a laboratory plasma experiment and initial results
which should give insight into the magnetic dynamics of accretion
discs and jets.  A high-speed multiple-frame CCD camera reveals images
of the formation and helical instability of a collimated plasma,
similar to MHD models of disc jets, and also plasma detachment
associated with spheromak formation, which may have relevance to disc
winds and flares.  The plasmas are produced by a planar magnetized
coaxial gun.  The resulting magnetic topology is dependent on the
details of magnetic helicity injection, namely the force-free state
eigenvalue $\agun$ imposed by the coaxial gun.
\end{abstract}


\section{Introduction}

The accretion disc occupies a leading role in astrophysics, figuring
prominently in young stellar objects (YSO), binary star systems, and
active galactic nuclei (AGN).  An unsolved mystery is the origin of
highly collimated bipolar jets and episodic flares associated with
accretion discs.  It was proposed some time ago that magnetic field
dynamics can supply the necessary jet formation and collimation
mechanisms (Blandford 1976; Lovelace 1976).   Magnetohydrodynamic
(MHD) simulations have shown that jets are a natural consequence of a
rotating disc in the presence of a magnetic field (\eg Shibata \&
Uchida 1985).  Jet structure, such as ``knots,''
and also episodic behavior are observed in the simulations (Ouyed \&
Pudritz 1997; Goodson, B\"{o}hm \& Winglee 1999; Nakamura, Uchida \&
Hirose 2001).  The details of these models are unlikely to be tested
by observations anytime in the near future.  Therefore, data from
laboratory experiments could be very useful for this purpose.  It
should be noted that astrophysical jets have been compared
theoretically with plasma guns (Contopoulos 1995), and that plasma
experimentalists have interpreted coaxial gun plasma flows in the
context of astrophysical jet morphology (Caress 1996).

This work describes a new plasma gun based laboratory experiment and
initial results which should give insight into the magnetic dynamics
of accretion discs and jets.  This experiment is the first to use a
plasma gun which explicitly simulates the geometry and topology of a
magnetically-linked star-disc system by using a co-planar disc-annulus
electrode setup.  The experiment reveals (1)~the formation and helical
structure of a magnetically-driven collimated plasma, similar to
proposed models of astrophysical jets, and (2)~plasma detachment which
may be relevant for disc winds and flares and field-line opening in
disc coronae. The resulting magnetic topology depends on the details
of magnetic helicity injection, namely the force-free state eigenvalue
$\alpha$ imposed at the boundary.  The plasmas in this work satisfy
the MHD criteria globally ($S\gg 1$, $\rhoi << L$, and $\VA << c$,
where $S$ is the Lundquist number, $\rhoi$ the ion gyro-radius, $L$
the plasma scale length, and $\VA$ the \Alfven speed).  It should be
noted that $S\sim 10^3$ in the experiment, similar to
that in MHD numerical simulations.

This paper is organized as follows.  Section~II gives a qualitative
description of how the magnetic dynamics of an accretion disc system
are modeled in the laboratory.  Section~III describes the experimental
setup.  Section~IV presents initial experimental results,
and Sec.~V provides a discussion.  Finally, a summary is given in Sec.~VI.

\section{Modeling disc magnetic dynamics in the laboratory}

For non-relativistic accretion disc dynamics, such as in YSO's and
some binary star systems, global magnetic field evolution can be
described using MHD.  The ideal MHD Ohm's law,
\begin{equation}
\Evec + \Uvec \times \Bvec = 0
\label{ohmslaw}
\end{equation}
implies that an electric field $\Evec$ will be induced due to the
sheared toroidal rotation of the disc in the presence of a magnetic
field $\Bvec$.  The $E_{\rm R}$ component creates an electric
potential drop $V$ between the central object and the accreting disc,
and therefore magnetic helicity $K$ will be injected into the disc
magnetosphere at the rate ${\rm d}K/{\rm d}t=2\psi V$ (\eg Bellan
2000), where $\psi$ is the poloidal magnetic flux emanating from the
central object.

In the laboratory, a planar coaxial disc-annulus electrode setup
simulates the topology and approximate geometry of the astrophysical
star-disc system.  However, rather than rotating the annulus to induce
an $E_{\rm R}$ as in a real accretion disc, an electric potential
$\Vgun$ is applied across the electrodes instead.  The background
poloidal $\Bvec$, which corresponds to the field of the central
object, is generated with an external coil.  Again through
Eq.~(\ref{ohmslaw}), the background $\Bvec$ and the applied voltage
induce an $\Evec \times \Bvec$ toroidal rotation in the plasma,
achieving a magnetic helicity injection rate ${\rm d}K/{\rm
d}t=2\psigun \Vgun$, where $\psigun$ is the poloidal flux intercepting
the inner disc electrode.  This setup is commonly used to create
spheromak plasmas (Rosenbluth \& Bussac 1979; Jarboe 1994; Bellan
2000).

Therefore, the most basic magnetic dynamics near a star-disc system,
namely the winding up of poloidal field lines generated by the central
object by disc rotation, can be simulated in the laboratory by
applying a voltage across coaxial electrodes in the presence of a
background poloidal field.  Of course there are limitations to the
laboratory setup.  For example, Keplerian sheared rotation and the
existence and placement of a star-disc co-rotation radius cannot be
exactly replicated.  An experiment cannot hope to capture all the
nuances and range of behaviors in all accretion disc systems.
However, it can provide experimental data on the most fundamental
plasma configurations which can result from a simple magnetically
driven system.

\section{Experimental setup}

\begin{figure}
\includegraphics[width=3.3truein]{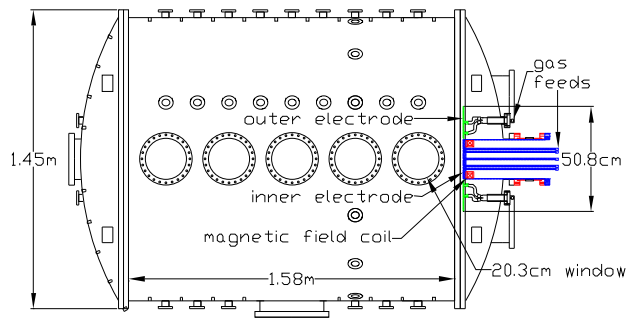}
\caption{Side-view schematic of vacuum chamber and planar coaxial gun
setup.}
\label{gun_chamber}
\end{figure}

\begin{figure}
\includegraphics[width=3.3truein]{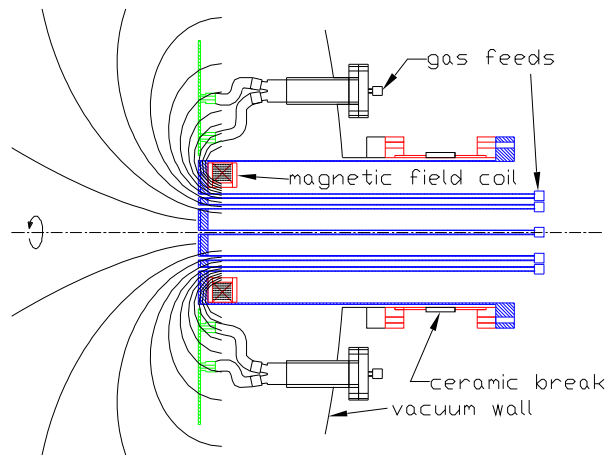}
\caption{Side-view schematic of planar coaxial gun, including gas
feeds, external magnetic field coil, poloidal flux $\psi$ contours,  
and symmetry axis.}
\label{gun_flux}
\end{figure}

Figure~\ref{gun_chamber} shows a side-view schematic of the
experimental setup.  A planar coaxial gun is installed  on one end of
a large vacuum chamber evacuated to $1.5 \times 10^{-7}$~Torr.  The
chamber is large compared to the plasma so as not to affect the plasma
evolution.  The gun setup, enlarged in Fig.~\ref{gun_flux}, includes
(1)~an inner electrode consisting of a 20.3~cm diameter copper disc
(attached to end of blue re-entrant port), (2)~an outer electrode
consisting of a 50.8~cm outer diameter copper annulus (in green),
(3)~an external solenoid (in red) to produce a poloidal bias magnetic
flux $\psi$ linking the inner and outer electrodes, and (4)~gas lines
to deliver fast puffs of neutral hydrogen gas to the desired position
of breakdown (adjacent to the electrodes).   The gap between
disc and annulus is 0.635~cm; plasma breakdown
does not occur in the gap because the pressure-distance product there
is too small to satisfy the Paschen breakdown criteria (Bellan 2000). 
A cylindrical coordinate
system ($R$, $\theta$, $Z$) is utilized, with the origin at the center
of the inner electrode and $+Z$ defined to be away from the electrode
(toward the left in Fig.~\ref{gun_flux}).  There are eight gas
injection holes on each electrode, distributed uniformly along the
$\theta$ direction of the inner and outer electrodes.  The bias field
is characterized by the parameter $\psigun$, which is the total
initial bias flux intercepted by the inner electrode; sample contours
of $\psi$ are shown in Fig.~\ref{gun_flux}.

The plasma formation sequence is as follows.  After the bias field and
gas puffs are introduced, negative polarity high-voltage is applied to
the inner electrode by discharging a 120$~\mu$F capacitor bank through
an ignitron; the outer electrode is maintained at vacuum chamber
ground.  In the presence of the high-voltage, the neutral hydrogen
breaks down.  The optimum path for breakdown is along vacuum field
lines linking the inner and outer electrodes as shown in
Fig.~\ref{gun_flux}.  The gun voltage $\Vgun$ drives a current $\Igun$
between the electrodes, twisting up the purely poloidal vacuum field
and producing a $\Bt$.  In an accretion disc, it is the disc rotation
which achieves the same effect.  The initally discrete flux tubes
expand and eventually merge into an axisymmetric configuration.
Throughout this process, magnetic helicity is being injected from the
gun into the plasma.

Experimental results to date have centered around time-resolved
global imaging of the plasma evolution.  Images shown here were taken using a
Cooke Corp.\ HSFC-PRO multiple-frame charge-coupled device (CCD)
camera, which takes up to eight images per plasma discharge.  The
camera view is through the 20.3~cm window (labeled in
Fig.~\ref{gun_chamber}) such that the gun electrodes appear on the
right hand side of each frame.  A false-color table is applied to the
images for ease of viewing.  Typically, the plasma is observed in
unfiltered visible light.  However, by using filters, it has been
verified that most of the light emission is from neutral hydrogen line
transitions.  The exposure time of each frame is 20~ns and the
interframe time is set typically to 1.5~$\mu$s, which is on the order
of an Alfv\'{e}n transit time.  Additionally, a triple Langmuir probe
is utilized to measure localized values of $\nee$, $\Te$, and floating
potential.  A Rogowski coil encircling the ceramic break (see
Fig.~\ref{gun_flux}) measures $\Igun$, and a high-voltage probe
measures $\Vgun$ at the mouth of the re-entrant port (shown in blue in
Fig.~\ref{gun_flux}).  Typical experimental parameters are as follows:
$V_{\rm gun} \approx 4$--6~kV, $I_{\rm gun} \approx 70$--130~kA,
$\psigun \approx 0.5$--2~mWb, $B \sim 0.2$--1~kG, $\Te \sim \Ti
\approx 5$--20~eV, $\nee \sim 10^{14}$~cm$^{-3}$, and global $\beta
\equiv 2\mu_0 nk(\Te+\Ti)/B^2 \sim 0.02$--$0.1$.

\section{Initial Experimental results}

Because the plasma is relatively low-$\beta$, it is reasonable as a
first approximation to consider it as nearly force-free with plasma
currents nearly parallel to the magnetic field, \ie
\begin{equation}
\nabla \times \Bvec \simeq \alpha \Bvec.
\label{ff-eq}
\end{equation}
Equation~(\ref{ff-eq}) implies that $\alpha$ is constant along field
lines but not necessarily across them.  Integrating Eq.~(\ref{ff-eq})
over the gun surface, it can be shown that the force-free state
eigenvalue imposed by the gun is $\agun = \mu_0 \Igun/\psigun.$
Experimentally, both $\Igun$ and $\psigun$ can be adjusted to achieve
a range of $\agun$ values.  Depending on the peak value of $\agun$,
several distinct plasma configurations are identified.

\subsection{Collimated plasma: an analog for disc jets}

\begin{figure*}
\begin{minipage}{6truein}
\includegraphics[width=6truein]{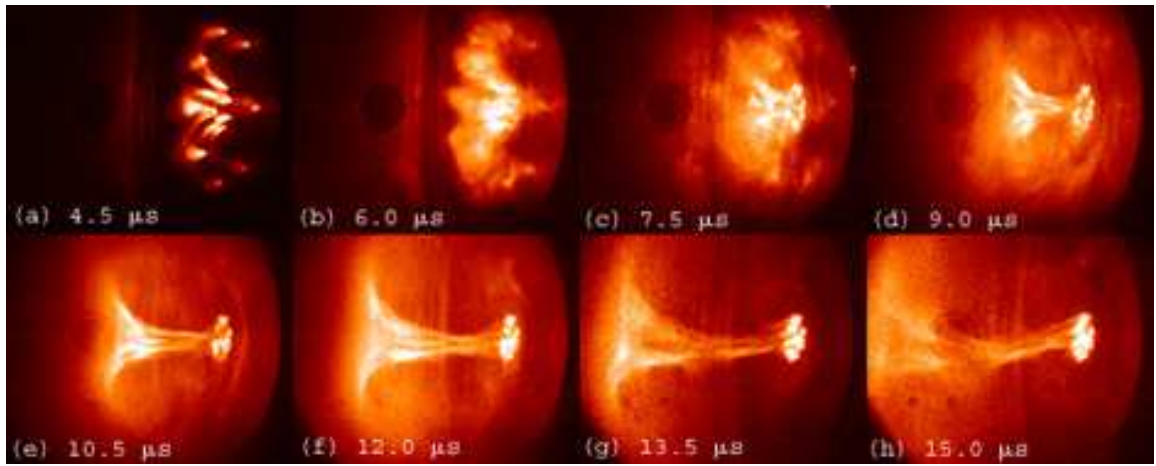}
\caption{Images of plasma evolution (shot 1210; peak $\agun \approx
66$~m$^{-1}$) in which a plasma column forms and persists for many
\Alfven transit times, illustrating the magnetic topology required for
an astrophysical jet.}
\label{1210_sequence}
\end{minipage}
\end{figure*}

For low values of $\agun$, the formation of a collimated plasma is
observed, as shown in Fig.~\ref{1210_sequence}. In this plasma,
$\agun$ peaks at approximately 66~m$^{-1}$.  In
Fig.~\ref{1210_sequence}(a), gas breakdown has just occurred along
eight discrete paths, each of which follows vacuum magnetic field
lines and terminates at gas injection holes.   In frame~(b), the
discrete arcs have expanded and begun coalescing.  Magnetic
reconnection is expected to occur along the $Z$-axis as the discrete
flux tubes coalesce; the particularly intense light emission there may
be due to higher plasma density resulting from compression.  In frames
(c)--(d), a central column forms and expands in the $Z$-direction.  In
frames~(e)--(g), the central column extends in length and persists for
several Alfv\'{e}n transit times.  In frame~(h), the central column
begins to break up.

The bright structures in the images are expected to correlate with
magnetic topology and current flow; this was verified previously in a
similar experiment using direct measurements of magnetic field (Yee \&
Bellan 2000).  This correlation is consistent with (1)~the light
emission being predominantly from neutral hydrogen atoms excited by
current-carrying electrons and (2)~the fact that the plasma is
low-$\beta$ and expected to be in a nearly force-free state in which
$\Jvec$ is nearly parallel to $\Bvec$.  It should be noted also that
H--H$^+$ charge-exchange time is estimated to be very fast ($\ll
1$~$\mu$s) in this experiment, and therefore light emission may be
representative of plasma ion dynamics.   Filamentary structures can be
seen inside the column, suggesting complex field topologies within the
globally collimated structure.

The cross section of the central column is clearly observable in
Fig.~\ref{1210_sequence}.  If there are no collimating forces, clearly
the length of the column would not be as elongated as observed, nor
would the cross section be so uniform  along the $Z$-direction.  This
result verifies that the necessary magnetic structure and collimation
for a disc jet can arise from magnetic forces associated with helicity
injection.

\subsection{Helical instability: an analog for jet structure}

\begin{figure*}
\begin{minipage}{6truein}
\includegraphics[width=6truein]{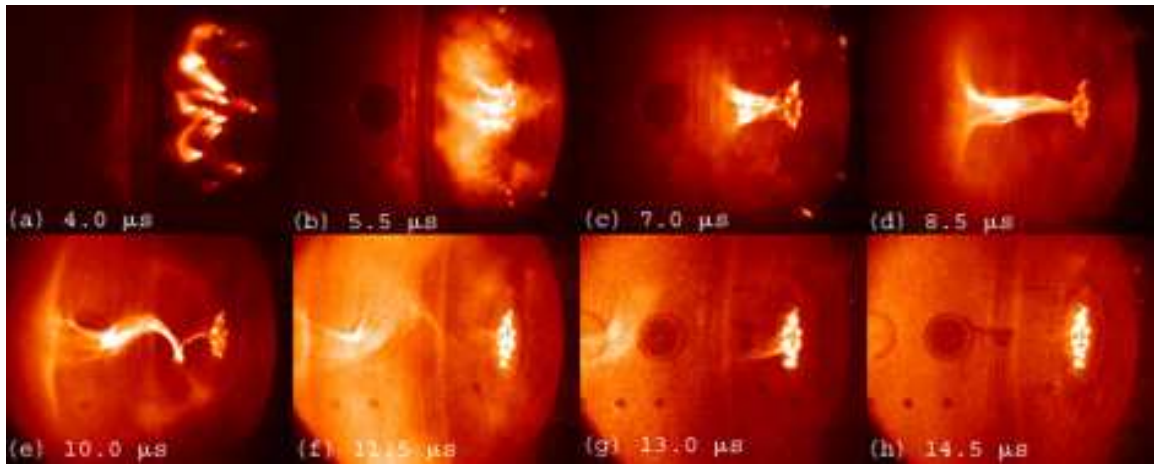}
\caption{Images of plasma evolution (shot 1233; peak $\agun \approx
71$~m$^{-1}$) in which a helical instability, likely a current-driven
kink, develops on the ideal MHD timescale, illustrating one possible
source of jet internal structure.}
\label{1233_sequence}
\end{minipage}
\end{figure*}

Higher $\agun$ results in plasmas with unstable central columns, as
shown in Fig.~\ref{1233_sequence}, in which $\agun$ peaks at
approximately 71~m$^{-1}$.  In Fig.~\ref{1233_sequence}(e), a highly
nonlinear helical perturbation appears in the collimated plasma.  By
altering the CCD camera timing, it is possible to determine the
characteristic growth time of the instability, which is about
1.7~$\mu$s, similar to the characteristic \Alfven time.  The
instability appears to be an ideal current-driven kink mode.

\begin{figure}
\includegraphics[width=3truein]{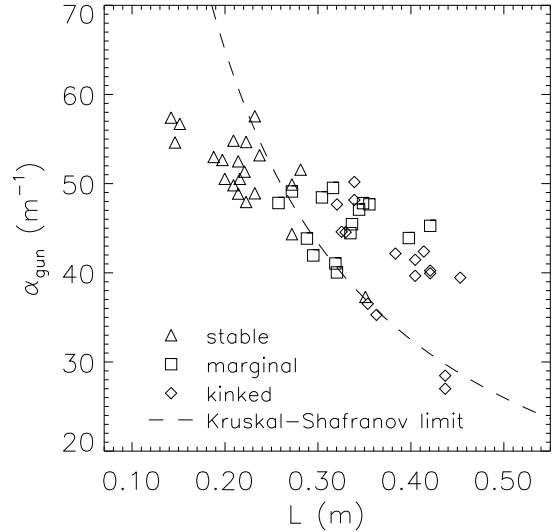}
\caption{A plot of $\agun$ versus column length $L$ for different
plasmas, showing good agreement with the Kruskal-Shafranov condition
for a current-driven kink instability.}
\label{KS}
\end{figure}

The Kruskal-Shafranov condition for ideal kink stability in a
cylindrical column is (\eg Bateman 1978)
\begin{equation}
q_{\rm edge} = 2\pi a \Bz/L\Bt > 1,
\end{equation}
where $a$ is the column radius and $L$ the column length.  Assuming
that both the initial $\psigun$ and the instantaneous $\Igun$ are
fully contained inside the plasma column, and using the definition of
$\agun$ and $\psigun$, it is straightforward to show that the
condition for stability is
\begin{equation}
\agun L < 4\pi.
\end{equation}
Figure~\ref{KS} plots $\agun$ versus $L$ for a collection of central
columns taken from both different discharges and different times from
within the same discharge.  Stable (triangles), marginally unstable
(squares), and kinked (diamonds) columns are plotted in $\agun$--$L$
space, along with the Kruskal-Shafranov stability threshold (dashed
line), showing good agreement between experiment and theory.  This
result indicates that plasma instabilities, in this case an ideal
kink, can give rise to macroscopic structure within jets.

\subsection{Detached plasma:  an analog for disc flares}

\begin{figure*}
\begin{minipage}{6truein}
\includegraphics[width=6truein]{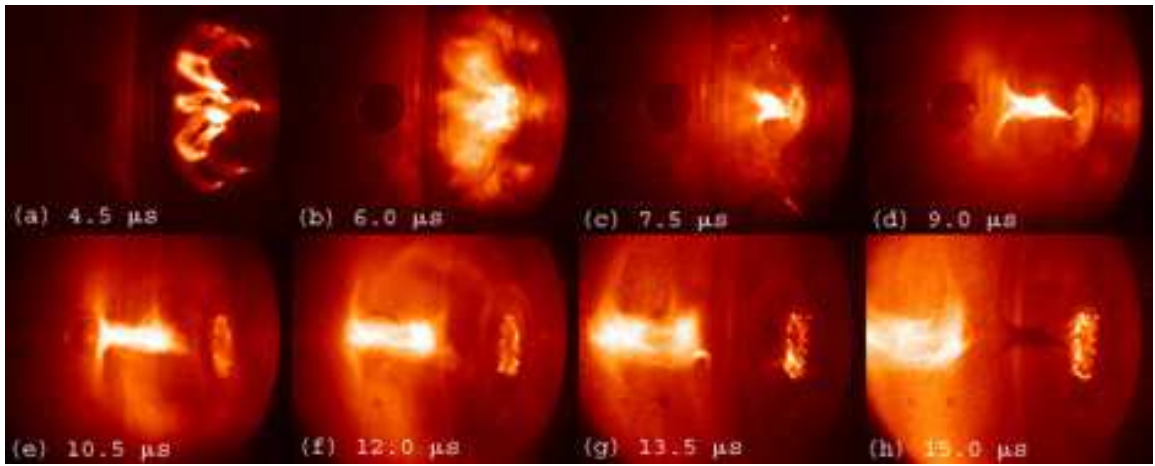}
\caption{Images of plasma evolution (shot 1181; peak $\agun \approx
129$~m$^{-1}$) in which the plasma detaches from the electrodes,
illustrating the possibility of field-line opening in disc coronae.}
\label{1181_sequence}
\end{minipage}
\end{figure*}

Still higher values of $\agun$ result in detached plasmas, as shown in
Fig.~\ref{1181_sequence}, in which $\agun$ peaks at approximately
129~m$^{-1}$.  In this discharge, the plasma appears to detach from
the electrodes in frame~(d) and then propagates along the $Z$
direction at a speed of approximately $6\times 10^4$~m/s, a fraction
of the estimated $\VA$.  It is likely that a spheromak configuration
is formed here; this was verified previously with direct measurement
of $\Bvec$ on an experiment with a non-planar source but similar
helicity injection (Yee \& Bellan 2000).  This result supports the
idea of field line reconnection above accretion discs, which can lead
to disc winds and also episodic high energy flaring.

\section{Discussion}

\begin{figure}
\includegraphics[width=3truein]{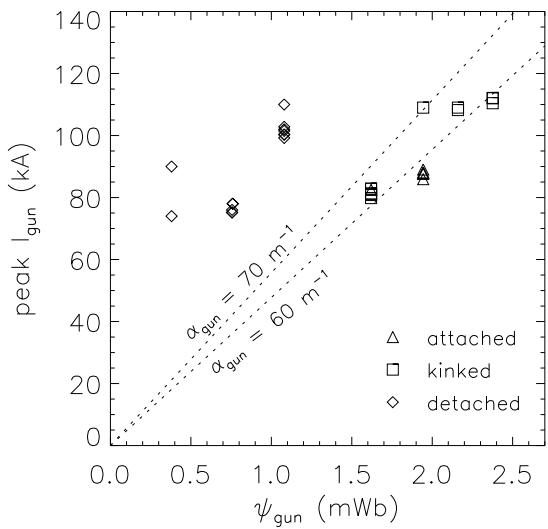}
\caption{A plot of peak $\Igun$ versus $\psigun$ for a collection of
plasma shots, showing the dependence of the plasma evolution on the
peak value $\agun = \mu_0 \Igun/\psigun$, with $\alpha_{\rm crit}
\approx 60$--70~m$^{-1}$.}
\label{lambda_plot}
\end{figure}

The results above show three distinct plasma configurations having
accretion disc characteristics.  All three configurations result from
the same plasma formation process, the only difference being the peak
value of the parameter $\agun = \mu_0 \Igun/\psigun$.
Figure~\ref{lambda_plot} illustrates this dependence by placing
different plasmas into $\Igun$--$\psigun$ parameter space, with
detachment at larger $\agun$, attached columns at lower $\agun$, and
kinked columns near $\alpha_{\rm crit}\approx 60$--70~m$^{-1}$.

This $\agun$ dependence has many implications.  Most importantly, it
suggests that the plasma configurations associated with accretion
discs and jets are related to Taylor relaxation theory (Taylor 1986),
which is a description of how a plasma evolves as magnetic energy is
minimized subject to the constraint of constant magnetic helicity.
This process can be cast as a variational problem, and it can be shown
that the resulting magnetic field configuration satisfies
Eq.~(\ref{ff-eq}) with uniform $\alpha$.  In the case of a driven
plasma like an astrophysical jet, $\nabla \alpha$ will be non-zero,
and magnetic helicity will flow from regions of high to low $\alpha$,
which tends to ``relax'' the plasma toward uniform $\alpha$ (Bellan
2000).  The evolution of many laboratory plasmas, including
spheromaks, can be understood in terms of Taylor relaxation.  For
example, it has been shown that a threshold $\alpha_{\rm crit}$ at the
source must be exceeded in order to form the closed-field
configuration of a spheromak (\eg Yee \& Bellan 2000).  This property
is not surprising since analytic solutions of Eq.~(\ref{ff-eq}) show
that $\Bvec$ transitions from sinh-like to sine functions above a
critical $\alpha$ (Bellan 2000).

Well-known models of astrophysical jets have considered separately the
roles played by poloidal (Blandford \& Payne 1982) and toroidal fields
(Contopoulos 1995) in jet formation.  However, the present work shows
that jet structure should result when the correct value of $\alpha$
occurs at the disc boundary, resulting in a non-arbitrary combination
of poloidal and toroidal fields in the disc corona.  Likewise,
deformation of the jet structure will occur under different choice of
$\alpha$.  In the experiment, the observed instability  is consistent
with an ideal current-driven kink.  This instability has been observed
in numerical simulations of AGN radio jets (Nakamura, Uchida \& Hirose
2001) and proposed as the responsible mechanism for wiggled structure
in the jets.

\section{Summary}

A plasma gun based laboratory experiment approximates the boundary
conditions and topology of a star-disc system, and it is shown
experimentally that magnetic helicity injection with these boundary
conditions leads naturally to both collimated plasmas and detached
plasmas, suggestive of disc jets and flares, respectively.  The onset
of a helical instability in the plasma column is shown to be
consistent with the Kruskal-Shafranov condition for an ideal
current-driven kink.  The magnetic topology depends on the force-free
state eigenvalue $\agun$ imposed at the plasma gun.  It is argued that
Taylor relaxation provides a useful description of the magnetic
structures of accretion discs and jets.  These results demonstrate
experimentally that the concept of magnetically-driven jets is a
viable one.  More quantitative characterization of the observed
plasmas is planned.  This will include direct measurements of the
magnetic field via insertable probes, ion flow velocity along the
collimated plasma via Doppler spectroscopy, and plasma pressure
profiles via Langmuir probes.

\section*{acknowledgments}

The authors are grateful to J.~Hansen, S.~Pracko,
C.~Romero-Talam\'{a}s, and F.~Cosso for technical assistance.  This
work was supported by a U.S. Dept.\ of Energy (US-DoE) Fusion Energy
Postdoctoral Fellowship and US-DoE contract no.\ DE-FG-03-98ER54461.

\bsp

\label{lastpage}
\end{document}